\documentclass[aps,prd,notitlepage,showpacs,nofootinbib,superscriptaddress]{revtex4-1}

\usepackage[utf8]{inputenc} 

\usepackage{amsmath}
\usepackage{amssymb}
\usepackage{float}
\usepackage{comment}
\usepackage{slashed}
\usepackage[normalem]{ulem}
\usepackage{bbold}
\usepackage{cancel}
\usepackage{graphicx}
\usepackage{multirow}
\usepackage{rotating}
\usepackage{caption}
\usepackage{subcaption}
\usepackage{tabulary}

\newcolumntype{K}[1]{>{\centering\arraybackslash}m{#1}}
\def\gsim{\raise0.3ex\hbox{$\;>$\kern-0.75em\raise-1.1ex\hbox{$\sim\;$}}}
\def\lsim{\raise0.3ex\hbox{$\;<$\kern-0.75em\raise-1.1ex\hbox{$\sim\;$}}}

\usepackage[usenames,dvipsnames]{color}

\newcommand {\ignore}[1]{}

\usepackage[colorinlistoftodos]{todonotes}
\usepackage[colorlinks=true,linkcolor=darkred,citecolor=darkred,urlcolor=darkred, pdfborder={0 0 0}]{hyperref}
\usepackage[normalem]{ulem}

\usepackage{caption}
\usepackage{subcaption}
\definecolor{linkcolor}{rgb}{0,0,0.5}

\definecolor{darkgreen}{rgb}{0,0.5,0}
\definecolor{darkred}{rgb}{0.6,0,0}
\definecolor{brown}{rgb}{0.59, 0.29, 0.0}
\definecolor{mightnightblue}{RGB}{25,25,112}

\def\vev#1{\left\langle #1\right\rangle}
\def\SM{$\mathrm{SU(3)_c \otimes SU(2)_L \otimes U(1)_Y}$ }

\def\Y{\mathbf{Y}}

\usepackage[force]{feynmp-auto}

\begin{document}


\title{\color{BrickRed} Dark-sector seeded solution to the strong CP problem}
\author{H. B. C\^amara}
\email{henrique.b.camara@tecnico.ulisboa.pt}
\affiliation{Departamento de F\'{\i}sica and CFTP, Instituto Superior T\'ecnico, Universidade de Lisboa, Lisboa, Portugal}
\author{F. R. Joaquim}
\email{filipe.joaquim@tecnico.ulisboa.pt}
\affiliation{Departamento de F\'{\i}sica and CFTP, Instituto Superior T\'ecnico, Universidade de Lisboa, Lisboa, Portugal}
\author{J.~W.~F. Valle}
\email{valle@ific.uv.es}
\affiliation{AHEP Group, Institut de F\'{i}sica Corpuscular --
  C.S.I.C./Universitat de Val\`{e}ncia, Parc Cient\'ific de Paterna.\\
 C/ Catedr\'atico Jos\'e Beltr\'an, 2 E-46980 Paterna (Valencia) - SPAIN}

\begin{abstract}
\vspace{0.5cm}
 We propose a novel realization of the Nelson-Barr mechanism ``seeded'' by a dark sector containing scalars and vector-like quarks. Charge-parity (CP) and a $\mathcal{Z}_8$ symmetry are spontaneously broken by the complex vacuum expectation value of a singlet scalar,
  leaving a residual $\mathcal{Z}_2$ symmetry that stabilizes dark matter~(DM). A complex Cabibbo-Kobayashi-Maskawa matrix arises via one-loop corrections to the quark mass matrix mediated by the dark sector. In contrast with other proposals where non-zero contributions to the strong CP phase arise at the one-loop level, in our case this only occurs at two loops, enhancing naturalness. Our scenario also provides a viable weakly interacting massive particle scalar DM candidate.
\end{abstract}

\maketitle
\noindent

\section{Introduction}

There are three experimental facts which call for new physics, beyond the Standard Model~(SM):
the observation of neutrino oscillations, the existence of some kind of dark matter (DM) and the matter-antimatter asymmetry of the Universe.
Besides these proofs of its incompleteness, some unaesthetic aspects of the SM also require a natural explanation.
One of these issues is the well-known strong charge-parity (CP) problem which can be formulated as a question:
why does quantum chromodynamics (QCD), the theory of strong interactions, seem to preserve CP when one would expect otherwise? 

CP violation (CPV) in QCD is encoded in the so-called strong CP phase $\bar{\theta}$ which induces nonvanishing contributions to the neutron electric dipole moment~(nEDM).
At present, the nEDM is constrained by experiment to be $\lesssim 3 \times 10^{-26} \; e \cdot \text{cm}$~\cite{Pendlebury:2015lrz,Baker:2006ts}, implying
\begin{equation}
\left|\overline{\theta}\right| \lesssim 10^{-10}\,.
\label{eq:thetabound}
\end{equation}
This seems to indicate that QCD does not violate CP at all. On the other hand, CP is maximally broken in weak interactions. From this point of view, tiny (or vanishing) CPV in the strong sector appears unnatural. A popular solution to the strong CP problem assumes a global anomalous Peccei-Quinn (PQ) symmetry which, after spontaneous breaking, gives rise to a pseudo-Goldstone boson --
the axion~\cite{Peccei:1977hh,Peccei:1977ur,Weinberg:1977ma}. 
The bottomline of the PQ mechanism is that axion dynamics leads to a CP-conserving ground-state, setting $\bar{\theta}=0$.

Another way of explaining the smallness of $\bar{\theta}$ is by simply imposing exact CP-symmetry at the Lagrangian level, ensuring a vanishing $\bar{\theta}$. 
However, to account for large CPV effects observed in the quark (weak) sector, CP must be broken spontaneously in such a way that low-energy CPV is large. This general setup~\cite{Nelson:1983zb,Nelson:1984hg,Barr:1984qx,Barr:1984fh} can be implemented in SM extensions with extra scalars and/or
colored particles which are crucial to break CP and generate a complex Cabbibo-Kobayashi-Maskawa (CKM) quark mixing matrix. The drawback of such Nelson-Barr (NB) type models is that, once CP is broken and the CP phase in the CKM matrix is large, quantum corrections to $\bar{\theta}$ must remain under control. 

The simplest way of accounting for spontaneous CP violation (SCPV) is by adding to the SM fields a complex scalar singlet $\sigma$ which acquires a vacuum expectation value (VEV). To transmit CPV to the SM quark sector, one may introduce a vectorlike quark~(VLQ) which couples to $\sigma$ in some way. Once CP is broken by the $\sigma$ VEV, CPV appears generating a complex CKM matrix. This is the essence of the model proposed by Bento, Branco and Parada (BBP) in Ref.~\cite{Bento:1991ez}. We note, however, that such minimal NB realization produces dangerous contributions to the strong CP phase already at the one-loop level,
thus requiring some rather strong assumptions to keep $\bar{\theta}$ under control, Eq.~\eqref{eq:thetabound}. 

Here we propose a new NB-type scenario, in which the strong CP phase arises only at two loops, while CPV in the CKM matrix arises via one-loop corrections mediated by a dark sector. After SCPV induced by the complex VEV of a scalar singlet $\sigma$, the dark particles remain odd under a $\mathcal{Z}_2$ symmetry, the lightest of them (a scalar) providing a viable weakly interacting massive particle~(WIMP) DM candidate. A key feature of our dark-mediated solution to the strong CP problem is that threshold corrections to $\overline{\theta}$ arise only at two-loops, alleviating the NB ``quality problem''.

\begin{table}[t!]
\renewcommand{\arraystretch}{1.4}
\setlength{\tabcolsep}{-1pt}
	\centering
	\begin{tabular}{| K{1.8cm} || K{1.5cm} | K{2.4cm} |  K{2.2cm} |}
		\hline 
 &Fields&$G_{\rm SM}$&  $\mathcal{Z}_8 \to \mathcal{Z}_2$ \\
		\hline \hline
		\multirow{6}{*}{Fermions}
&$q_L$&($\mathbf{3},\mathbf{2}, {1/6}$)& {$\omega^2$}   $\to$  $+$  \\
&$u_R$&($\mathbf{3},\mathbf{1}, {2/3}$)& {$\omega^2$}   $\to$  $+$  \\
&$d_{R}$&($\mathbf{3},\mathbf{1}, {-1/3}$)& {$\omega^2$}   $\to$  $+$  \\
&$B_{L,R}$&($\mathbf{3},\mathbf{1}, {-1/3}$)& {$\omega^6$}   $\to$  $+$  \\
&$D_{1 L,1R}$&($\mathbf{3},\mathbf{1}, {-1/3}$)& {$\omega^7$}   $\to$  $-$  \\
&$D_{2 L,2R}$&($\mathbf{3},\mathbf{1}, {-1/3}$)& {$\omega^3$}   $\to$  $-$  \\
		\hline \hline
		\multirow{4}{*}{Scalars}
&$\Phi$&($\mathbf{1},\mathbf{2}, {1/2}$)&{$1$}    $\to$ $+$ \\	
&$\sigma$&($\mathbf{1},\mathbf{1}, {0}$)&{$\omega^2$}  $\to$  $+$ \\
&$\chi$&($\mathbf{1},\mathbf{1}, {0}$)&{$\omega^3$}    $\to$ $-$ \\	
&$\xi$&($\mathbf{1},\mathbf{1}, {0}$)&{$\omega$}  $\to$  $-$ \\
\hline
	\end{tabular}
	\caption{Field content and their transformation properties under the SM gauge and $\mathcal{Z}_8$ symmetries, where $\omega^k = e^{i\pi k/4}$, and under the remnant $\mathcal{Z}_2$ after spontaneous $\mathcal{Z}_8$ breaking.}
	\label{tab:model}
\end{table}
\section{Model at tree level}
A crucial ingredient in our construction is SCPV, which is simply realized by the VEV of a complex scalar singlet $\sigma$. 
This is possible if the scalar potential of the theory includes phase-sensitive terms as, e.g., $\sigma^4$ and $\sigma^2$,
invariant under a $\mathcal{Z}_N$ discrete symmetry if $\sigma \to \omega^{k}$ with $\omega=e^{2i\pi /N}$ and $k=pN/4\,(p \in \mathbb{Z})$. 
Our minimal choice is $N=8$. Thus, besides gauge invariance under the SM group G$_{\rm SM}=$ \SM and under CP, our theory also has a $\mathcal{Z}_8$ symmetry.

To implement our dark-matter-mediated NB solution to the strong CP problem, we add three down-type VLQs, namely one VLQ $B_{L,R}$, and two odd (dark) $D_{i L,iR}$ ($i=1,2$). Besides $\sigma$ and the SM Higgs doublet $\Phi$, we also have two inert complex scalar singlets $\chi$ and $\xi$, which are also dark. The transformation properties of all fields under G$_{\rm SM}$ and the $\mathcal{Z}_8$ symmetry are shown in Table~\ref{tab:model}.  
We denote the SM left-handed quark doublets and right-handed up/down quark singlets by $q_L=(u_L\; d_L)^T$ and $u_R/d_R$, respectively, with Yukawa interactions
\begin{align}
- \mathcal{L}_{\text{Yuk}} &\supset \Y_{u}  \overline{q_L} \tilde{\Phi} u_R + \Y_{d}  \overline{q_L} \Phi d_R + \Y_\xi  \overline{D_{2 L}} d_R \xi + \Y_\chi \;  \overline{D_{1 L}} d_R \chi^\ast+ \text{H.c.} \;,
\label{eq:LYukSM}
\end{align}
where $\Phi = \left( \phi^+ \; \phi^0 \right)^T$ and $\tilde{\Phi}=i\tau_2 \Phi^\ast$, $\tau_2$ being the complex Pauli matrix. Here $\Y_{u,d}$ ($\Y_{\chi,\xi}$) are $3 \times 3$ ($1 \times 3$) matrices and, as usual, $\vev{\phi^0}=v/\sqrt{2} \simeq 174$~GeV. The Yukawa couplings involving only new fields read
\begin{align}
- \mathcal{L}_{\text{Yuk}} &\supset  y_\chi \;  \overline{B_L} D_{2 R} \chi + y_\xi \;  \overline{B_L} D_{1 R} \xi^\ast + y_\chi^\prime \;  \overline{D_{2 L}} B_R \chi^\ast + y_\xi^\prime \;  \overline{D_{1 L}} B_R \xi + \text{H.c.} \;,
\label{eq:LYukchi}
\end{align}
where $y_{\chi,\xi}^{(\prime)}$ are numbers, and bare VLQ mass terms are
\begin{align}
- \mathcal{L}_{\text{mass}} &= m_B \; \overline{B_L} B_R + m_{D_{1,2}} \; \overline{D_{1,2 L}} D_{1,2 R}  + \text{H.c.} \,.
\label{eq:Lbare}
\end{align}
Notice that Eqs.~\eqref{eq:LYukSM}-\eqref{eq:Lbare} contain all gauge-invariant Yukawa and mass terms which respect the $\mathcal{Z}_8$ symmetry.
CP invariance of the Lagrangian implies that all coupling and mass parameters are real.

The $\mathcal{Z}_8$ symmetry is broken down to a $\mathcal{Z}_2$ (see Table~\ref{tab:model}) by the $\sigma$ VEV $\vev{ \sigma}= v_\sigma\, e^{i \varphi}/\sqrt{2}$. In the limit of exact $\mathcal{Z}_8$ invariance, the only phase-sensitive term in the scalar potential is $\lambda_\sigma(\sigma^4+\sigma^{*4})$. Minimization leads to $\varphi=\pi/4+k\pi/2\,(k \in \mathbb{Z})$. Note that this solution does not violate CP, since a generalized CP transformation can be defined such that the vacuum remains invariant. Furthermore, spontaneous breaking of an exact discrete symmetry could lead to cosmological domain-wall problems~\footnote{This might not be an issue if our mechanism is embedded in a more general framework providing a solution to that problem~(see e.g. \cite{McNamara:2022lrw}).}. We, thus, consider a scenario in which the $\mathcal{Z}_8$ is softly broken by the bilinear term $m_\sigma^{2}(\sigma^2+\sigma^{*2})$, fixing the domain-wall problem. This leads to a CP-violating phase $\varphi$ that can, in principle, be arbitrary.
 
It is straightforward to see that, since there are no $\mathcal{Z}_8$-invariant quark-$\sigma$ couplings, the $4\times 4$ tree-level down-quark mass matrix
$\mathcal{M}_d^{(0)}$ in the $(d\;B)_{L,R}$ basis is block-diagonal and real, with the SM quarks decoupled from the VLQ $B$. Hence CPV will not be communicated to the quark sector and the CKM matrix is real \footnote{In contrast, in Ref.~\cite{Bento:1991ez} the allowed couplings $\overline{B}_L d_R \sigma^{(\ast)}$ would yield a complex
$\overline{B}_L d_R$ mass term and a complex tree-level CKM.}. Since
\begin{equation}
  \bar{\theta} = \arg[\det(\mathbf{M}_u)] + \arg[\det(\mathcal{M}_d)]\,, 
\end{equation}
where $\mathbf{M}_u=\Y_u v/2$ is the SM up-quark mass matrix, we obviously have $\bar{\theta} =0$.

\section{Complex CKM at one loop with $\mathbf{\bar{\theta}=0}$} 
Beyond tree-level, the down-quark mass matrix can be written in the generic form $\mathcal{M}_d = \mathcal{M}_d ^{(0)}+ \Delta \mathcal{M}_d$ with
\begin{equation}
  \mathcal{M}_d^{(0)} =\begin{pmatrix} \mathbf{M}_d  & 0 \\ 0 & m_B \end{pmatrix}, 
  \Delta \mathcal{M}_d =
  \begin{pmatrix} \Delta\mathbf{M}_d  & \Delta \mathbf{M}_{d B} \\ \Delta \mathbf{M}_{Bd} & \Delta m_B \end{pmatrix},
\label{eq:Lmassloopcorrection}
\end{equation}
where $\mathbf{M}_d=\Y_d v/\sqrt{2}$ and $m_B$ is the bare $B$ mass term, see Eqs.~\eqref{eq:LYukSM} and \eqref{eq:Lbare}. Higher-order corrections to $\mathcal{M}_d ^{(0)}$ are encoded in $\Delta\mathcal{M}_d$ and a necessary condition to generate a complex effective CKM matrix
is that at least one of the correcting terms is complex. The most intuitive way of investigating how this may happen is to look for higher-order operators which can generate complex mass terms after SCPV. Such operators must be gauge and $\mathcal{Z}_8$ invariant and contain unmatched powers of $\sigma^{(\ast)}$, given that the $\sigma$ VEV phase $\varphi$ is the only source of CPV in our framework. Then one must check at which loop order those operators arise and compute the corresponding corrections $\Delta\mathcal{M}_d$. 

At dimension five, phase-sensitive operators which induce corrections to $\mathcal{M}_d$ are $\overline{B_L} d_R  \sigma^{(\ast)2}$, these specifically contribute, after SCPV, to $\Delta \mathbf{M}_{Bd}$. In contrast, the operators $\overline{B_L} B_R \; (\Phi^\dagger \Phi)$ and $\overline{B_L} B_R \; |\sigma|^2$ lead to real $\Delta m_B$. Notice that, since $\sigma$ does not couple to quarks, we require interactions with the dark sector to induce those operators at the quantum level.

\begin{figure}[t!]
\centering
    \includegraphics[scale=0.86]{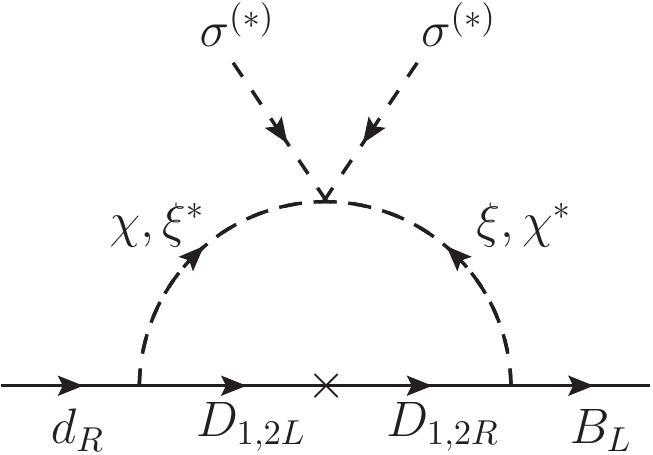}  \hspace{+1.5cm} \includegraphics[scale=0.83]{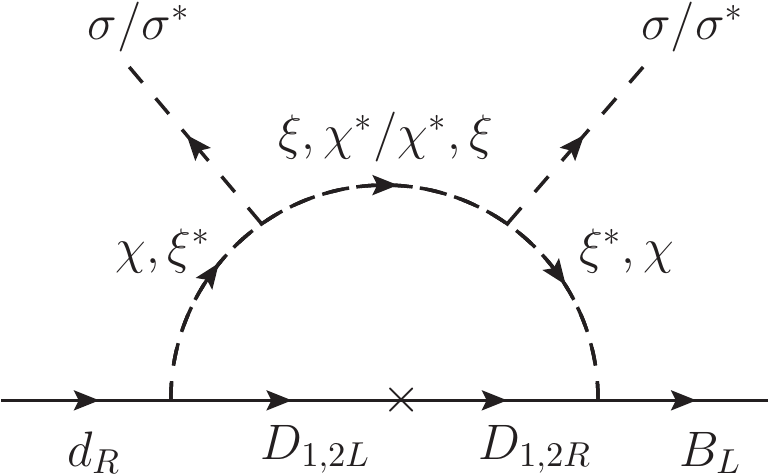}
    \caption{``Dark-mediated'' diagrams for the dim-5 operators $\overline{B_L} d_R\sigma^{(*)2}$ leading to
    $\Delta \mathbf{M}_{Bd}$ after $\mathcal{Z}_8$ symmetry breaking.}
    \label{fig:oneloopdiags}
\end{figure}
 The lowest-order phase-sensitive operators induced at one-loop are $\overline{B_L} d_R  \sigma^{(\ast) 2}$, which generate $\Delta\mathbf{M}_{Bd}$ after symmetry breaking. The corresponding Feynman diagrams in the weak basis are shown in Fig.~\ref{fig:oneloopdiags}. The trilinear and quartic scalar terms involving $\sigma$ and the $dark$ fields $\zeta=\chi,\xi$ are all $\mathcal{Z}_8$ symmetric. The contributions in Fig.~\ref{fig:oneloopdiags}, are roughly estimated as:
\begin{align}
|\Delta\mathbf{M}_{Bd}| &\sim \frac{1}{16 \pi^2}  \lambda_{\sigma \zeta\zeta} |\mathbf{Y}_\zeta| \,y_\zeta \frac{v_\sigma^2}{m_\zeta^2}\,m_D \; ,\label{eq:MBestimation1}\\
|\Delta\mathbf{M}_{Bd}| &\sim \frac{1}{16 \pi^2}   |\mathbf{Y}_\zeta| \,y_\zeta \,\frac{\mu_\zeta^2}{m_\zeta^2} \frac{v_\sigma^2}{m_{\zeta}^2} \,m_D\,,
    \label{eq:MBestimation2}
\end{align}
for the left and right diagram, respectively. Here $\mathbf{Y}_\zeta$ and $y_\zeta$ represent generic $\mathbf{Y}_{\chi,\xi}$ and $y_{\chi,\xi}$ couplings of Eq.~\eqref{eq:LYukchi}, while $\lambda_{\sigma \zeta\zeta}$ and $\mu_{\zeta}$ are quartic and trilinear terms of the scalar potential. It is clear that $\Delta\mathbf{M}_{Bd}$ is complex due to the interference of different terms which pick up the phases $\pm 2\varphi$ from the VEVs of $\sigma^2$ and $\sigma^{*2}$. Similar one-loop diagrams exist for $\overline{B_L} B_R \; (\Phi^\dagger \Phi)$ and $\overline{B_L} B_R \; |\sigma|^2$, these however lead to a real $\Delta m_B$. 

The one-loop down-quark mass matrix is then:
\begin{equation}
  \mathcal{M}_d^{(1)} =\begin{pmatrix} \mathbf{M}_d  & 0 \\  \Delta\mathbf{M}_{Bd} & \widehat{m}_B \end{pmatrix} \,,\, \widehat{m}_B=m_B+\Delta m_B
   \, .
\label{eq:Md1loop}
\end{equation}
In the limit $\mathbf{M}_d \ll  \widehat{m}_B$, the (complex) CKM matrix can be obtained diagonalizing $\mathbf{M}_{\text{light}}^2$ given by
\begin{align}
\mathbf{M}_{\text{light}}^2 \simeq \mathbf{M}_d \mathbf{M}_d^T - \frac{\mathbf{M}_d \Delta \mathbf{M}_{Bd}^\dagger \Delta\mathbf{M}_{Bd} \mathbf{M}_d^T}{\widetilde{m}_B^2} \, , \label{eq:Mlight} 
\end{align}
with $\widetilde{m}_B^2 \simeq |\Delta \mathbf{M}_{Bd}|^2 + \widehat{m}_B^2$. Whether CPV is successfully transmitted to the SM sector depends on the relative size between $\Delta \mathbf{M}_{Bd}$ and $\widehat{m}_B$.
In fact, in this case, generating a viable CKM requires  $|\Delta \mathbf{M}_{Bd}| \gtrsim \widehat{m}_B$. 

Notice that, $\bar{\theta} = \arg[\det(\mathbf{M}_u)] + \arg[\det(\mathcal{M}_d)]=0$, since $\mathbf{M}_d$ and $\widehat{m}_B$ are real, and $\Delta \mathbf{M}_{d B}=0$. This is the key feature of our dark-seeded NB mechanism, which is in contrast with the BBP model where corrections to $\bar{\theta}$ appear already at
the one-loop level. In our case, $\bar{\theta}$ remains zero at this order of perturbation theory.

\section{Corrections beyond one loop}
At the two-loop level, complex corrections to $ \mathbf{M}_d$ and $m_B$ induce contributions to $\bar{\theta}$ which can be estimated as
    \begin{align}
     \Delta \overline{\theta}|_{\Delta \mathbf{M}_{d}}&\sim \frac{1}{(16 \pi^2)^2} \; \lambda_{\Phi \sigma}\, y_d^2 \,\frac{v_\sigma^2}{v^2}   \,, \\
    \Delta \overline{\theta}|_{\Delta m_B} & \sim \frac{1}{(16 \pi^2)^2} \, \lambda_{\sigma \zeta} \ y_\zeta \,y_\zeta^\prime \ \frac{m_D}{m_B} \frac{v_\sigma^2}{m_\zeta^2}  \label{eq:2loopdmb}\,,
    \end{align}
    where $m_\zeta$ is a typical dark scalar mass, and $y_\zeta^{(\prime)}$ are generic $y_{\xi,\chi}^{(\prime)}$ couplings. Here $\lambda_{\Phi \sigma}$ is the $\left(\Phi^\dagger \Phi\right) |\sigma|^2$ quartic scalar coupling and $\lambda_{\sigma \zeta}$ stands for generic
    $\lambda_{\sigma \chi} |\sigma|^2 |\chi|^2$ and $\lambda_{\sigma \xi} |\sigma|^2 |\xi|^2$ couplings. For typical values for the SM quark Yukawa couplings $y_d \sim \mathcal{O}(10^{-2})$, the first correction above is under control if
    $\lambda_{\Phi \sigma} \lesssim v^2/v_\sigma^2$. This is reasonable, as the physics accounting for the Higgs hierarchy is likely to also provide a small $\lambda_{\Phi \sigma}$. On the other hand, if all mass scales in Eq.~\eqref{eq:2loopdmb} are of the same order, $\Delta \overline{\theta}|_{\Delta m_B} \lesssim 10^{-10}$
    requires $|\lambda_{\sigma \zeta} \ y_\zeta \,y_\zeta^\prime| \lesssim 10^{-6}$, which can be easily accommodated. In fact, in our framework, the U(1)-sensitive couplings with the dark sector can be naturally small in the 't Hooft sense~\cite{tHooft:1979rat} since the Lagrangian symmetry is enlarged in their absence. Note that, the above contributions come from operators $\overline{q_L}\Phi d_R \sigma^{(\ast) 4}$ and $\overline{B_L} B_R \sigma^{(\ast) 4}$.

    Concerning higher-loop corrections, we have checked that the contributions to $\overline{\theta}$ arise from three (four) loops via $\Delta\mathbf{M}_{d,dB}$ ($m_B$), which can be estimated as
\begin{align}
     \Delta \overline{\theta}|_{\Delta \mathbf{M}_{d B}} &\sim \frac{\Delta \overline{\theta}|_{\Delta \mathbf{M}_{d}}}{16\pi^2}\sim \frac{1}{(16 \pi^2)^2} \; \lambda_{\Phi \sigma} \frac{|\Delta\mathbf{M}_{Bd}|^2}{v_\sigma^2}   \; , \\
    \Delta \overline{\theta}|_{\Delta m_B} & \sim \frac{g^2}{(16 \pi^2)^2} \; \frac{|\Delta\mathbf{M}_{Bd}|^2}{v^2_\sigma} \; ,
\end{align}
where $g \sim \mathcal{O}(1)$ is a weak coupling and we have considered a $\mathcal{O}(1)$ coupling for the $|\sigma|^4$ term. It is straightforward to see that $|\Delta\mathbf{M}_{Bd}| \lesssim 10^{-3} v_\sigma$ is required to keep these corrections under control
(as long as $\lambda_{\Phi \sigma}$ is made small in a framework where the Higgs mass is stabilized). One may now ask how natural is it to verify this condition in our scenario. In the above estimates, $\Delta\mathbf{M}_{Bd}$ is the one-loop correction in Eq.~\eqref{eq:Md1loop}, see Fig.~\ref{fig:oneloopdiags} and Eqs.~\eqref{eq:MBestimation1} and~\eqref{eq:MBestimation2}. From those estimates one sees that, to ensure $|\Delta\mathbf{M}_{Bd}| \lesssim 10^{-3} v_\sigma$ one roughly needs
$ |\mathbf{Y}_\zeta| \,y_\zeta \lesssim  m_\zeta^2/(m_D v_\sigma)$ for $\lambda_{\sigma\zeta\zeta} \lesssim 1$ and $ \mu_\zeta\sim m_\zeta$. This condition is attainable for reasonable values of dark sector couplings and wide mass ranges. In contrast, models where $\mathbf{M}_{Bd}$ is generated at tree-level via a $y_B\sigma^{(\ast)} \overline{B_L}d_R$ have been argued to suffer from a ``quality problem'',
requiring a small $y_B \lesssim 10^{-3}$~\cite{Perez:2020dbw,Valenti:2021rdu}. 

Indeed, in the original BBP scenario,  $\Delta \mathbf{M}_{d B}$ and $\Delta m_B$ receive contributions from dim-5 operators of the type
$\overline{q_L}\Phi B_R \sigma^{(\ast)}$ and $\overline{B_L} B_R \sigma^{(\ast)2}$, respectively. These affect $\overline{\theta}$ in a way that $\Delta \overline{\theta} \lesssim 10^{-10}$ sets an upper bound on the SCPV scale $v_\sigma \lesssim 10^{3} - 10^{8}$~GeV,
for a cutoff $\Lambda$ at the Planck scale~\cite{Choi:1992xp,Dine:2015jga}. This hierarchy between $v_\sigma$ and $\Lambda$ is the essence of the NB ``quality problem''. 
As recently noted in ref.~\cite{Asadi:2022vys}, such low SCPV scale may have a drastic impact in cosmology. In our case, the lowest dimension operators that would induce corrections to $\overline{\theta}$ are the dim-6 $y_\Lambda\overline{q_L} \Phi B_R \sigma^{(*)2}$, for which we estimate
\begin{equation}
    \Delta \overline{\theta}|_{\Delta \mathbf{M}_{d B}} \sim \frac{|\Delta\mathbf{M}_{Bd}|}{m_B} \frac{y_\Lambda}{y_d}  \left( \frac{v_\sigma}{\Lambda} \right)^2  \; .
\end{equation}
Taking $|\Delta\mathbf{M}_{Bd}|/m_B \gtrsim \mathcal{O}(1)$ to generate a viable complex CKM matrix, and $y_\Lambda \sim \mathcal{O}(1)$ with $y_d \sim 10^{-5}-1$,
we get that $\Delta \overline{\theta}|_{\Delta \mathbf{M}_{d B}} \lesssim 10^{-10}$ only requires $v_\sigma \lesssim 10^{8} - 10^{13}$~GeV, a milder hierarchy between those scales.

\begin{figure}[t!]
    \centering 
    \begin{tabular}{ll}
            \includegraphics[scale=0.45]{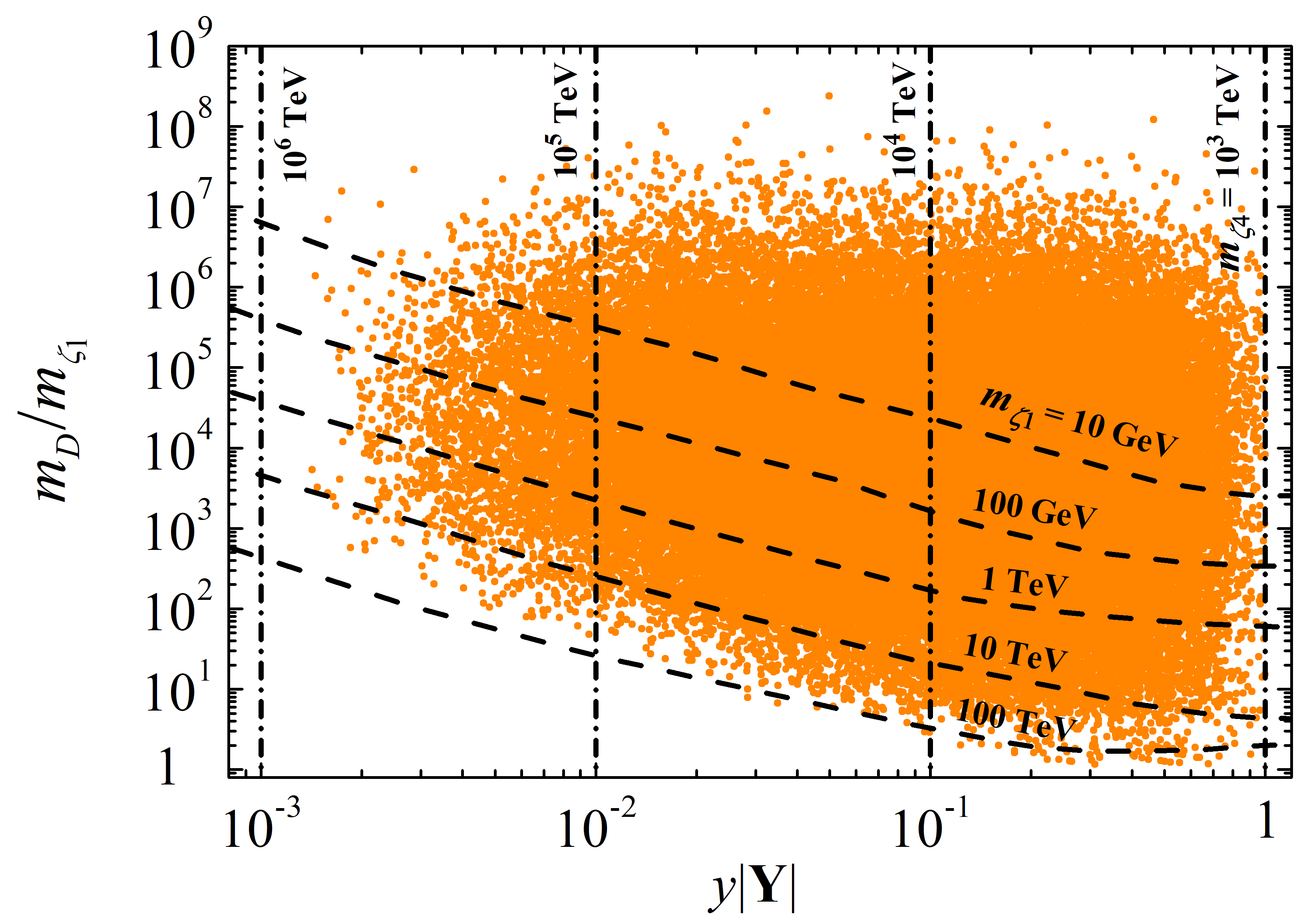}
    \end{tabular}
    \caption{$m_D/m_{\zeta_1}$ versus $y |\mathbf{Y}|$, where $m_D = (m_{D_1} + m_{D_2})/2$ and $y |\mathbf{Y}| = (y_\chi |\mathbf{Y}_\xi| + y_\xi |\mathbf{Y}_\chi|)/2$ -- see Eqs.~\eqref{eq:LYukSM}-\eqref{eq:Lbare}. We set $v_\sigma=10^3$~TeV. Above the dashed contours, $m_{\zeta_1}$ lies below the labelled value.  The same holds to the right of the dash-dotted vertical lines for the heaviest dark-scalar mass $m_{\zeta_4}$.}  
    \label{fig:CKMDM}
 \end{figure}

\section{Phenomenology}
We have seen that $|\Delta\mathbf{M}_{Bd}| \lesssim 10^{-3} v_\sigma$ and $|\Delta\mathbf{M}_{Bd}| \gtrsim \widehat{m}_B$ are needed to simultaneously satisfy
the $\bar{\theta}$ bound of Eq.~\eqref{eq:thetabound} and successfully transmit CPV to the CKM matrix. These constraints, together with the 1.4~TeV LHC limit on the $B$ VLQ mass~\cite{CMS:2020ttz}, imply $v_\sigma \gtrsim 10^3$~TeV. Fig.~\ref{fig:CKMDM} shows a scatter plot of $m_D/m_{\zeta_1}$ versus $|y|\Y|$, with quark masses and CKM parameters within their $1\sigma$ experimental
ranges~\cite{ParticleDataGroup:2022pth}, and the $B$ VLQ mass above the LHC limit. All results have been obtained using exact one-loop computation of $\Delta\mathbf{M}_{Bd}$ and diagonalizing the full $\mathcal{M}_d^{(1)}$. Notice we obtain viable points over a wide range of dark couplings and masses.

\begin{figure}[t!]
    \centering
    \hspace{-1cm} \includegraphics[scale=0.47]{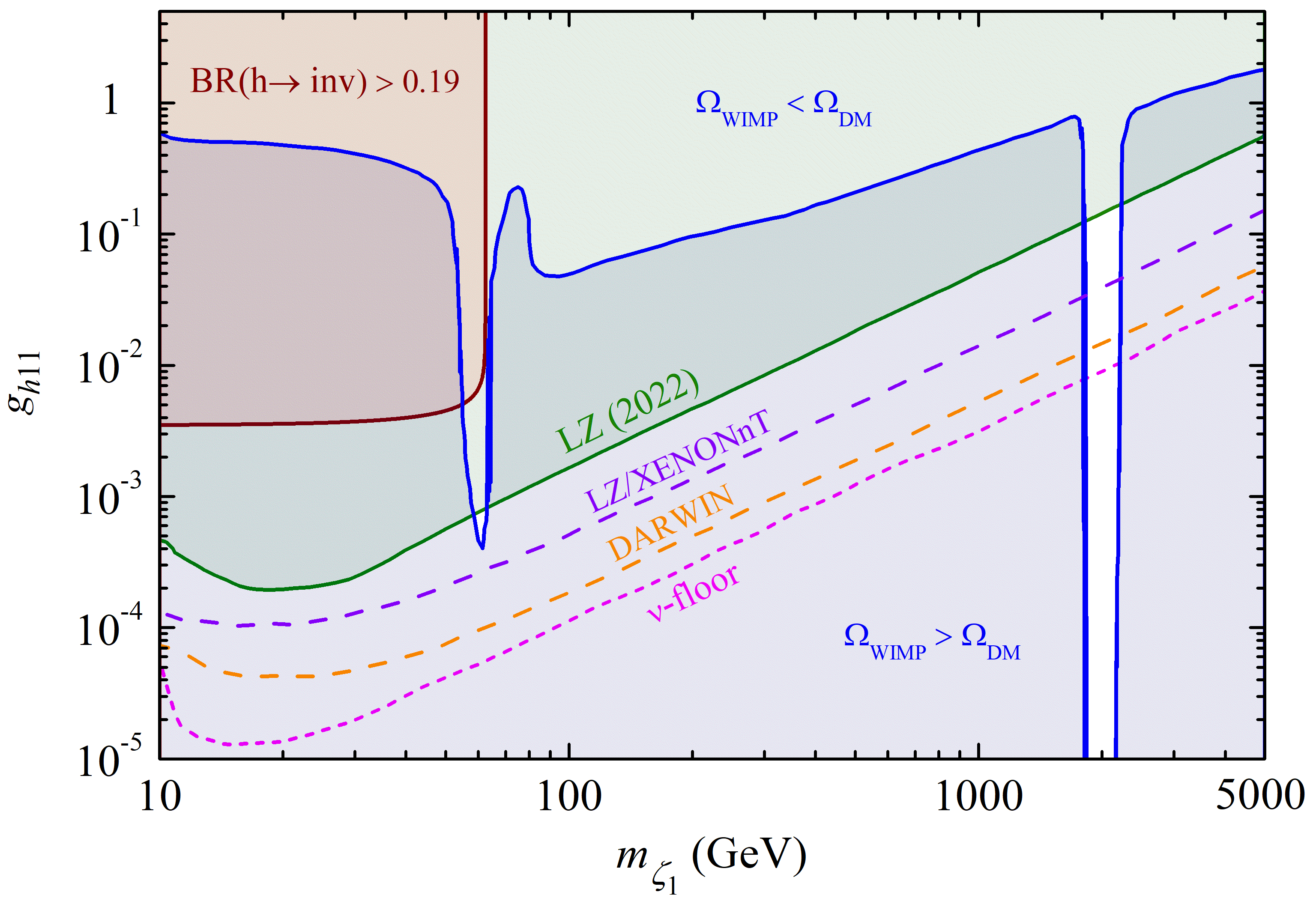}
    \caption{Higgs-DM coupling $g_{h 1 1}$ versus WIMP DM mass $m_{\zeta_1}$. Along the blue contour the DM relic density lies in the Planck $3\sigma$ range~\cite{Planck:2018vyg}. The blue shaded region below that leads to overabundant DM. The green shaded region is excluded by the LZ experiment~\cite{LZ:2022ufs}. The violet and orange contours indicate the projected sensitivities for LZ~\cite{LZ:2018qzl}, XENONnT~\cite{XENON:2020kmp}, and DARWIN~\cite{DARWIN:2016hyl}, respectively. The pink dashed line is the "neutrino floor" limit~\cite{Billard:2013qya}. The brown-shaded region is excluded by the LHC bound on the Higgs invisible decay~\cite{ParticleDataGroup:2022pth}.}
    \label{fig:DMplot}
  \end{figure}

Concerning DM, we assume a benchmark dark scalar mass spectrum of the type $m_{\zeta_1} \ll m_{\zeta_{2,3,4}}$, being $\zeta_1$ our DM candidate. As seen in Fig.~\ref{fig:DMplot} our scenario differs from the simplest scalar-singlet DM case~\cite{McDonald:1993ex,Guo:2010hq,Cline:2013gha,Feng:2014vea,GAMBIT:2017gge,Casas:2017jjg}
due to the presence of even scalars $H_{1,2}$ arising from $\sigma$. Besides the viable relic density dip at $m_{\zeta_1} \sim m_h/2 \simeq 62.6$~GeV (SM Higgs boson), $H_{1,2}$ open up new annihilation channels which reproduce the observed DM relic abundance. As shown in the figure, our dark-sector can be probed by future direct detection experiments, e.g. LZ~\cite{LZ:2018qzl}, XENONnT~\cite{XENON:2020kmp}, and DARWIN~\cite{DARWIN:2016hyl}.

\section{Concluding remarks}
In this paper, we propose a new solution to the strong CP problem based on the existence of a dark sector containing a viable (scalar) WIMP DM candidate, as seen in Fig.~\ref{fig:DMplot}. In our NB-inspired mechanism, a $\mathcal{Z}_8$ symmetry allows for SCPV while leaving a residual $\mathcal{Z}_2$ to stabilize DM. A complex CKM matrix arises from one-loop corrections to the quark mass matrix mediated by the dark sector; see Figs.~\ref{fig:oneloopdiags} and~\ref{fig:CKMDM}. In contrast with other proposals, here the strong CP phase receives non-zero contributions only at two loops, enhancing naturalness.

Our setup can be embedded in a more general framework aiming at addressing other drawbacks of the SM, besides the strong CP problem and DM. For instance, the VEV of the complex scalar singlet $\sigma$ could be responsible for generating neutrino masses, inducing simultaneously low-energy CP violation in the lepton mixing matrix~\cite{Barreiros:2020gxu}. Moreover, the same scalar may also play a key role in creating the lepton asymmetry required for leptogenesis~\cite{Barreiros:2022fpi} as well as driving inflation~\cite{Boucenna:2014uma}. This opens a window for interesting studies where a dark sector provides a unique solution to several open questions in (astro)particle physics and cosmology.

\begin{acknowledgments}
This work is supported by Funda\c{c}\~ao para a Ci\^encia e a Tecnologia (FCT, Portugal), projects CFTP-FCT Unit UIDB/00777/2020, UIDP/00777/2020,
and CERN/FIS-PAR/0019/2021, partially funded by POCTI (FEDER), COMPETE, QREN and EU, 
and also by the Spanish grants PID2020-113775GB-I00 (AEI/10.13039/501100011033) and Prometeo CIPROM/2021/054 (Generalitat Valenciana).
H.B.C. is supported by the PhD FCT grant 2021.06340.BD.
\end{acknowledgments}

\end{document}